# Raman spectra of epitaxial graphene on SiC and of epitaxial graphene transferred to SiO$_2$

*Dong Su Lee,[*] Christian Riedl, Benjamin Krauß, Klaus von Klitzing, Ulrich Starke, and Jurgen H. Smet*

Max-Planck-Institut für Festkörperforschung, Heisenbergstr. 1, D-70569 Stuttgart, Germany

Raman spectra were measured for mono-, bi- and trilayer graphene grown on SiC by solid state graphitization, whereby the number of layers was pre-assigned by angle-resolved ultraviolet photoemission spectroscopy. It was found that the only unambiguous fingerprint in Raman spectroscopy to identify the number of layers for graphene on SiC(0001) is the linewidth of the 2D (or D*) peak. The Raman spectra of epitaxial graphene show significant differences as compared to micromechanically cleaved graphene obtained from highly oriented pyrolytic graphite crystals. The G peak is found to be blue-shifted. The 2D peak does not exhibit any obvious shoulder structures but it is much broader and almost resembles a single-peak even for multilayers. Flakes of epitaxial graphene were transferred from SiC onto SiO$_2$ for further Raman studies. A comparison of the Raman data obtained for graphene on SiC with data for epitaxial graphene transferred to SiO$_2$ reveals that the G peak blue-shift is clearly due to the SiC substrate. The broadened 2D peak however stems from the graphene structure itself and not from the substrate.

Graphene is the building block of graphite and carbon-based nanomaterials such as carbon nanotubes and fullerenes. Since the development of the micromechanical cleavage method to obtain thermodynamically stable mono- and few-layer graphene, our understanding of this purely two-dimensional system has improved significantly.[1,2,3,4] Graphene exhibits unconventional electronic properties,[1,2,5,6] such as high and nearly equal mobilities at room temperature for both electron and hole conduction, which makes it a strong candidate for nanoelectronic circuit applications.[5,7] However, the micromechanical cleavage method is not suitable for obtaining large area graphene. For practical applications requiring large areas of graphene, full graphitization on SiC seems to be the more promising route.[8,9,10,11] Indeed, devices of epitaxial graphene on SiC have been prepared using conventional e-beam lithography and a patterning method based on $O_2$ plasma etching[8,9] although several problems still remain: the nature of the $(6\sqrt{3}\times 6\sqrt{3})R30°$ reconstruction at the interface between SiC and graphene is still under debate and the proper conditions for the production of large areas of homogeneous mono-, bi- and few-layer graphene are not well developed.[10,12]

Raman spectroscopy is known to be a powerful tool to determine the electronic properties of carbon-based materials and there have been several reports about Raman measurements on graphene layers which were micromechanically exfoliated from highly oriented pyrolytic graphite (HOPG) on $SiO_2$ substrates.[13,14,15,16,17] Recently, Ferrari et al. have demonstrated that the shape of the 2D Raman peak may serve as the fingerprint to distinguish mono-, bi- and few-layer graphene. The 2D peak stems from a double resonance electron-phonon scattering process.[14] For monolayer graphene the 2D peak can be fitted to a single Lorentzian, whereas the multiple bands in bilayers or few-layer graphene require fitting to 4 or more Lorentzians. Although Raman data of few-layer epitaxial graphene were recently reported,[18,19,20] a similar, clear procedure to differentiate between single layer, bilayer and multilayer graphene on SiC is lacking. In this letter, we have identified a clear Raman fingerprint to distinguish the layer thickness by correlating Raman spectroscopy experiments with ultraviolet photoemission spectroscopy on mono-, bi-, tri- and multilayer epitaxial graphene. In addition we show Raman data of epitaxial graphene flakes transferred onto $SiO_2$ in order to investigate the influence of the substrate material or buffer layer on the Raman spectrum.

The graphene preparation was conducted by graphitization of n-type 4H- and 6H-SiC samples in ultra high vacuum (UHV).[10,21,22] Before loading the samples into the UHV chamber, they were hydrogen etched in order to remove polishing damage. By applying this procedure it is possible to obtain a regular array of atomically flat terraces of approximately 2 µm width. Moreover, this treatment chemically passivates the surface.[23] The samples were prepared in UHV by Si deposition (1ML/min) and annealing[24] in order to get a well defined starting point for the graphene preparation. For the growth of the graphene layers in a second heating step, temperatures between 1200 °C and 1600 °C were chosen, depending on the desired degree of graphitization.[10,25] For a series of SiC(0001) samples the number of layers was determined *in situ* by angle-resolved ultraviolet photoemission spectroscopy (ARUPS) and/or low energy electron diffraction (LEED)[25] before they were moved to the Raman apparatus. For an additional series of samples, the graphene layer thicknesses were determined from an optical technique described in the Supporting Information. This series includes graphene samples on both SiC(0001), the so-called Si-face or Si-terminated SiC, and SiC(000$\bar{1}$) referred to as the C-face or C-terminated SiC. To transfer graphene flakes from SiC onto Si substrates with a 300 nm thick thermal oxide the conventional adhesive tape approach was used. Details are discussed in the Supporting Information.

Raman spectra were measured under ambient conditions using an $Ar^+$ laser with a wavelength of 488 nm. The incident laser power was 12 mW, the spot size ~ 1 µm in diameter and the exposure time 30 min. Because the Raman signal of graphene is much smaller than that of SiC, as we will discuss later, it is necessary to use high power levels and long acquisition times. It was verified that no significant spectral changes were induced by the extensive laser exposure during Raman scans with long acquisition times.

The studied mono- and bilayer graphene samples were grown on 6H-SiC(0001), the trilayer graphene samples were prepared on 4H-SiC(0001). Figures 1a-c show ARUPS color renditions of the energy versus momentum as well as a cross section through the data at a binding energy, $E_B = 0.8$ eV. Using these data sets, the number of layers of the graphene samples can be unambiguously determined.[25] It should be noted that it was shown recently, that LEED spot intensities can also be used as a fingerprint for the number of graphene layers on SiC(0001).[25] The area probed by ARUPS and LEED amounts to ~ 1 mm$^2$. The Fermi energy ($E_F$) is shifted from the Dirac crossing point ($E_D$) by ~ 430, 310 and 300 meV for mono-, bi- and trilayer graphene, respectively. The energy positions of the Dirac point are marked by yellow lines in Fig. 1. These shifts are consistent with previous calculations[26] as well as experiments,[11,27] where the $E_F$ shift was in addition found to be independent of the polytypes or the doping level of the SiC substrate. The shifts are, however, influenced by the number of layers. Hence, the electron doping is believed to be associated with surface charges at the interface with the buffer layer.[27] The conical electronic band near the K point is described by a Dirac dispersion and the charge concentration $n$ for monolayer graphene can be derived as $|n| = (E_F - E_D)^2 / (\pi \hbar^2 v_F^2)$, where $h/2\pi$ is Planck's constant, $v_F$ the Fermi velocity ($v_F = 1.1 \times 10^6$ m/s)$^2$. The resulting electron density is about $n \sim 1 \times 10^{13}$ cm$^{-2}$ for monolayer graphene. The carrier concentration changes, however, after the samples are taken out of the UHV chamber. ARUPS measurements for monolayer graphene after exposure to air show a Dirac energy of 260 meV below the Fermi level (not shown). Using this value, we obtain an electron concentration of $n \sim 4.2 \times 10^{12}$ cm$^{-2}$. The reduced electron concentration may be due to compensation by p-type dopants such as adsorbed oxygen or water molecules from the ambient environment.

The key Raman results are displayed in Fig. 1d. Spectra are plotted for a monolayer, a bilayer and a triple layer of graphene on SiC(0001), together with reference spectra measured on bare 4H-SiC and 6H-SiC substrates as well as on bulk HOPG and mechanically exfoliated monolayer and bilayer graphene from HOPG which were transferred onto an oxidized Si substrate. The well isolated G and 2D peaks for HOPG and exfoliated graphene layers on SiO$_2$ near 1580 cm$^{-1}$ and 2700 cm$^{-1}$ jump into the eye. The defect D peak, which should be near 1350 cm$^{-1}$ is absent for these layers attesting the high crystalline quality of graphene layers exfoliated from HOPG. Note also that the 2D peaks of the monolayer and bilayer appear different. For the bilayer a shoulder structure is clearly visible. For the epitaxial graphene layers on SiC(0001), it is apparently difficult to distinguish the G and D peaks in the spectra since their intensities are much smaller than those of the surrounding peaks from the SiC substrate. This can be concluded from the reference spectra which were recorded on pure SiC samples immediately after hydrogen etching. Only the 2D peaks are well isolated around ~ 2700 – 2750 cm$^{-1}$ and are out of the range of the SiC peaks. Higher resolution measurements on epitaxial graphene as well as the SiC reference samples using a grating with more lines are depicted in Fig. 2. In these spectra, the D and G peaks can be discerned more easily. Boxes mark the approximate location of the D and G-features. But even now it is clearly difficult to obtain quantitative information about the G and D peaks in order to identify the layer thickness from the raw data as plotted in Fig. 2a-b. Fig. 2c shows differential spectra obtained by subtracting the Raman spectrum of the SiC reference sample from the spectra of the graphitized samples. This procedure has recently been introduced by Röhrl et al.[20] Now, a small D peak is visible near ~ 1360 cm$^{-1}$. Also the G peak, located in between the SiC peaks at ~ 1500 and ~ 1620 cm$^{-1}$, is brought out in these differential spectra. The G peak position (P(G)) does not vary with the number of layers (P(G) ~ 1591 cm$^{-1}$). However, it is blue-shifted as compared to micromechanically cleaved graphene (P(G) ~ 1587 cm$^{-1}$) and graphite (see Fig. 1d). The full width at half maximum (FWHM) of the G peak (FWHM(G)) is also not influenced by the number of layers.

The number of layers for graphene samples exfoliated from HOPG has been successfully assigned by analyzing the shape of the 2D peak.[14] Exfoliated monolayer graphene for instance exhibits a single and sharp 2D peak (FWHM < 30 cm$^{-1}$) that can be fitted to a single Lorentzian. For bilayer and trilayer graphene the 2D feature becomes broader (50 cm$^{-1}$ for the bilayer in Fig. 1c for instance), asymmetric and shoulder structures appear. Fitting requires multiple Lorentzians, four in the case of a bilayer. In the following, we demonstrate that it is not straightforward to determine the number of layers in the case of epitaxial graphene by simple visual inspection of the shape of the 2D peak in the raw data. For our epitaxially grown graphene, the 2D peak position, P(2D) increases monotonously from 2721 to 2760 cm$^{-1}$ with increasing layer number (Fig. 2d). The peak widths FWHM(2D) are substantially larger than for exfoliated layers and rise from ~ 46 to 64 and 74 cm$^{-1}$ when comparing a monolayer sample with bi- and trilayer graphene, respectively. Despite the broadening, the 2D peak of monolayer epitaxial graphene can still be fitted with a single Lorentzian (Fig. 2(e)). In bilayer graphene grown on SiC, however, visual inspection of the 2D peak in Fig. 2d shows no clear asymmetry, nor a shoulder structure. Even if so, the peak can only be poorly fitted by a single Lorentzian as demonstrated in Fig. 2f. As in the case of exfoliated bilayers, a decomposition into four Lorentzians is necessary to obtain better agreement between the fit and the experimental data. Even for up to 14 stacked layers of graphene the shape of the 2D peak does not develop a shoulder structure but rather resembles closely a single peak in contrast to exfoliated layers from HOPG. This observation is in agreement with previously reported Raman data for epitaxial graphene.[18, 19]

The P(2D) and FWHM(2D) values are plotted as a function of the inverse number of layers in Fig. 3. The stars represent the mono-, bi- and trilayer data discussed above. The number of layers for the remaining data points is determined either using LEED intensities[25] or an optical attenuation model based on Beer's Law (details can be found in the Supporting Information).[19,28] The tendency of P(2D) as the number of layers increases is not unambiguous. There is a relatively large spread of P(2D) even for samples with the same number of layers. Hence, P(2D) is not suitable to precisely identify the number of graphene layers a sample has. However, for graphene on SiC(0001), the FWHM(2D) value exhibits a clear linear relationship with the inverse number of layers: $FHWM(2D) = (-45(1/N) + 88)$ [cm$^{-1}$]. Note that for samples prepared on SiC(000$\bar{1}$), the FWHM(2D) does not vary with layer thickness. This suggests that the electronic properties of graphene layers prepared on SiC(000$\bar{1}$) are quite different from those of samples fabricated on SiC(0001).[8,26,29] In summary, the width of the 2D peak stands out as a suitable fingerprint to assign the number of layers in the case of graphene on SiC(0001). It is possible to analyze the peak shape (single peak or four peak contributions) to distinguish monolayer from bilayer graphene provided a high resolution spectrum is recorded. Since the 2D peak amplitudes are small and since the shape of the 2D peak for bilayer graphene can visually not be distinguished from a single peak, long acquisition times are needed for a reliable shape analysis.

We now return to the discrepancies in the G peak position (blue-shift) and the width of the 2D peak for graphene on SiC compared with graphene exfoliated from HOPG. The following issues may be relevant: epitaxial graphene layers may have different defects and curvature and the as grown graphene layers likely interact strongly with the $(6\sqrt{3}\times6\sqrt{3})R30°$ buffer layer reconstruction. To shed some light on the role of this buffer layer for the Raman features, we compare Raman data of epitaxial graphene on the SiC substrate and expitaxial graphene which was transferred onto a Si substrate with a 300 nm thick $SiO_2$ top layer. The transferred graphene flakes on $SiO_2$ were typically about 1 $\mu m^2$ or less in size due to the initial morphology of the graphene layers on SiC.[10,25,30,31] With the help of atomic force microscopy (AFM) two different types of graphene flakes can be distinguished (Fig. 4a - c). Some flakes show a flat surface over their entire area similar to normal graphene flakes exfoliated from HOPG (Fig. 4a), but more often flakes are not flat as illustrated in Fig. 4b and c. This suggests that – even after the transfer onto $SiO_2$ – the flakes might maintain some of their initial morphology imposed by the fine terrace structure of the SiC substrate formed after the annealing process.[10] Although the flakes were not flat the Raman data for different points on each flake were nearly the same. Fig. 4d depicts a representative Raman spectrum obtained on a graphene monolayer exfoliated from SiC onto $SiO_2$. Raman data around the 2D peak are displayed in Fig. 4e for a monolayer, bilayer and a multilayer (more than 2) sample. Finally, the position of the 2D peak as well as the FWHM of the G and 2D features is shown in Fig. 4f and g for a large collection of flakes exfoliated from epitaxial graphene. The shape of the 2D peak changes upon the transfer of the flakes and allows to determine the layer thickness for each flake by analyzing the shape of the 2D peak together with the ratio of the G and 2D peak strength, I(G)/I(2D).[13] Accordingly, the flakes are classified into three groups: monolayer (black), bilayer (red) and multilayer (more than 2, blue).[14] For all flakes, the D peak intensity is small as shown in the inset of Fig. 4f around half of the flakes have an I(D)/I(G) ratio of less than 0.05), which indicates high crystalline quality of the initial epitaxial graphene on SiC, since the disorder would certainly not be reduced after the graphene flakes are transferred.

The G peak is red-shifted after transferring epitaxial graphene onto $SiO_2$, ~ 1587 ± 2 $cm^{-1}$ back to the position for exfoliated graphene from HOPG. Also very similar, the exact location of the G peak appears to be randomly distributed without any correlation to the number of layers (not shown) and also its peak width is more or less constant (Fig. 4g). Hence, the G peak blue-shift for graphene on SiC can be attributed to the presence of the SiC substrate and buffer layer. A possible cause for this blue-shift may be compressive strain which builds up during the cool down procedure[20] or charge doping from the substrate.[16, 32,33,34,35] Assuming that charge neutrality exists at ~ 1583 cm$^{-1}$ (the minimum value of the measured P(G) for the flakes), the G peak blue-shift of ~ 8 cm$^{-1}$ from 1583 to 1591 cm$^{-1}$ (the value for graphene on SiC) corresponds to $n \sim 5 \times 10^{12}$ cm$^{-2}$ according to the literature.[33,35] This carrier density is consistent with the carrier density estimated from ARUPS data ($n \sim 4.2 \times 10^{12}$ cm$^{-2}$, see above). For the 2D peak, there is no significant red-shift after the graphene has been transferred to SiO$_2$ (cf. Fig. 4f). This suggests that the SiC substrate does not cause phonon stiffening which would influence the 2D peak. However, in view of the large variation of P(2D) a definite conclusion can not yet be drawn.

It has been previously reported that the 2D peak for epitaxial graphene on both SiC(0001)[19] and SiC(000$\bar{1}$)[18,19] does not exhibit a clear asymmetry or shoulder structure. For SiC(000$\bar{1}$) this was attributed to the "turbostratic" stacking which decouples the layers electronically.[8,18,29] We have grown up to 14 graphene layers on our SiC(0001) samples and still obtained a 2D Raman line with the shape of a single peak. Such thick layers are most likely *AB*-stacked. Upon transferring to the SiO$_2$ substrate, the 2D peak becomes more asymmetric as seen in Fig. 4e. The peak now largely resembles that of exfoliated multilayer graphene from HOPG. Hence, the anomalous shape of the 2D peak of epitaxial graphene on SiC cannot be assigned solely to the stacking sequence, at least for graphene on the Si-terminated surface of SiC(0001), because the stacking should not change after being transferred onto SiO$_2$. Instead, presumably either the strain from the (6√3×6√3)R30° reconstruction or charge doping affect the double resonance process.

Although the shape of the 2D peak changes after the graphene flakes are transferred, the peak width does not change as shown in Fig. 4g. The FWHM(2D) value for the monolayer graphene flakes is ~ 46 ± 4 cm$^{-1}$, which is similar to the value of graphene on SiC. Thus, apparently the broadening of the 2D peak of graphene on SiC does not originate from a substrate effect but is due to the graphene structure itself. One possible explanation for the broadening of the 2D peak is a non-uniform number of layers: if a monolayer graphene sample on SiC would consist of a small fraction of bi- or multilayer, the 2D peak would be broadened[17] whereas the effect would be smeared out in the ARUPS measurement on the large scale of ~ 1 mm$^2$. However, this explanation can be disproved by confocal Raman maps using a diffraction limited laser spot size of 400 nm on different positions of the monolayer graphene grown on SiC. We find that the terraces themselves do not have a large variation of P(2D) and FWHM(2D) and so the number of layers is quite homogeneous (see Supporting Information). We propose, that the intrinsic disorder in epitaxial graphene and notably the curvature introduced near the numerous small steps in the morphology formed during the graphitization process is responsible for the 2D peak broadening.

In summary, we have measured Raman spectra for mono-, bi-, tri- and multilayer graphene grown on SiC after the number of layers was determined by ARUPS. From the analysis of the Raman peaks we find that FWHM(2D) stands out as the unambiguous fingerprint to assign the number of layers for graphene samples fabricated on SiC(0001). Graphene flakes prepared on SiC were successfully transferred onto SiO$_2$. Comparing the Raman data recorded for graphene on SiC, exfoliated graphene from SiC onto SiO$_2$ and micromechanically cleaved graphene from HOPG on SiO$_2$, we demonstrated that the G peak is blue-shifted. This shift and the single-peak appearance of the 2D feature of epitaxial graphene on SiC are caused by the SiC substrate. The observed broadening of the 2D peak cannot be attributed to the SiC substrate but is likely caused by the morphology which develops in the course of the graphitization procedure.

ACKNOWLEDGMENT

D.S.L. was supported by the Korea Research Foundation Grant funded by the Korean Government (MOEHRD):KRF-2007-357-C00027.

Available Supporting Information: Discussion on the morphology of epitaxial graphene on SiC, determination of the graphene film thickness using an attenuation model, details on transferring epitaxial graphene from SiC to SiO$_2$ and spatially resolved Raman measurements on epitaxial monolayer graphene. This material is available free of charge via the Internet at http://pubs.acs.org.

FIGURE CAPTIONS

Figure 1**a**–**c** ARUPS images and cross-section (momentum distribution curve, MDC) at $E_B = 0.8$ eV of mono- (**a**), bi- (**b**) and trilayer graphene (**c**) grown on SiC(0001). The arrows in the MDC data (upper panel) indicate the position of the π-bands. Yellow lines mark the Dirac energy $E_D$. **d**: Raman spectra of HOPG, mono- and bilayer graphene exfoliated from HOPG and transferred onto $SiO_2$, mono-, bi- and trilayer graphene grown on SiC (monolayer, bilayer: 6H-SiC(0001), trilayer: 4H-SiC(0001)) and reference data on bare 4H-SiC and 6H-SiC substrate pieces after hydrogen etching. All data were obtained using a wavelength of 488 nm. The scale for the data of the graphene samples on SiC and bare SiC chips are normalized with respect to the highest SiC peak ~ 1510 cm$^{-1}$. The data of HOPG and the graphene flakes from HOPG are scaled down arbitrarily to fit the image.

Figure 2**a**–**b** Raman spectra around the D and G peaks of (**a**) mono- (black) and bilayer graphene (red) together with the 6H-SiC(0001) reference (green) and of (**b**) trilayer graphene (blue) with the 4H-SiC(0001) reference (green). The dotted boxes mark the vicinity of the D and G peaks, which are partly masked by Raman contributions from the SiC substrate. **c**: Differential spectra between the Raman data recorded on the graphene samples and on the bare SiC reference pieces. **d**: Raman spectra near the 2D peak for epitaxial graphene samples with 1, 2 and 3 layers. The data are normalized to the peak height. **e**–**f**: Curve fitting (yellow lines) of the spectra around the 2D peak, for monolayer graphene (filled circles in e) with a single Lorentzian and for bilayer graphene (filled circles in f) with four Lorentzian curves. The lower panel shows the bilayer spectrum (filled circles) and a best fit to the data using a single Lorentzian curve.

Figure 3. P(2D) and FWHM(2D) of graphene grown on SiC samples as a function of the inverse number of layers. The different symbols refer to the different methods used to identify the layer thickness. Stars represent data obtained on mono-, bi- and trilayer graphene, which were grown on SiC(0001) and whose layer thickness was determined via *in-situ* ARUPS. Filled diamonds are the data where the number of layers was assigned by LEED measurements. The number of layers of the other samples (filled squares and open circles) was estimated by calculating the attenuation of the laser light as described in the Supporting Information. The open circles mark samples grown on SiC(000$\bar{1}$). All other samples (filled symbols) were fabricated on SiC(0001). The color of the data points refers to different polytypes of SiC: black for 6H-SiC and red for 4H-SiC. The dashed line in the lower panel is a least square linear fit through the data of graphene samples grown on the Si-face of SiC (stars and squares).

Figure 4. **a**–**c**: AFM images of representative graphene flakes transferred from SiC onto $SiO_2$. All scale bars are 500 nm. **d**: A representative Raman spectrum recorded on a monolayer graphene flake. **e**: Raman spectra around the 2D peak for transferred mono-, bi- and multilayer flakes. **f**: P(2D) as a function of I(G)/I(2D) of the graphene flakes. Black circles correspond to monolayer samples, red squares to bilayer and blue triangles to multilayer samples. The inset depicts a histogram of I(D)/I(G) of the graphene flakes. **g**: FWHM(2D) and FWHM(G) as a function of I(G)/I(2D). The FWHM(2D) increases but the FWHM(G) decreases slightly with the number of layers. The dashed lines are guides to the eye.

REFERENCES

[1] Novoselov, K. S.; Geim, A. K.; Morozov, S. V.; Jiang, D.; Zhang, Y.; Dubonos, S. V.; Grigorieva, I. V.; Firsov, A. A. Science, **2004**, 306, 666 – 669

[2] Novoselov, K. S.; Geim, A. K.; Morozov, S. V.; Jiang, D.; Katsnelson, M. I.; Grigorieva, I. V.; Dubonos, S. V.; Firsov, A. A. Nature, **2005**, 438, 197 – 200.

[3] Zhang, Y.; Tan, Y.-W.; Stormer, H. L.; Kim, P. Nature, **2005**, 438, 201 – 204.

[4] Meyer, J. C.; Geim, A. K.; Katsnelson, M. I.; Novoselov, K. S.; Booth, T. J.; Roth, S. Nature, **2007**, 446, 60 – 63.

[5] Geim, A. K.; Novoselov, K. S. Nat. Mater. **2007**, 6, 183 – 191.

[6] Semenoff, G. W. Phys. Rev. Lett. **1984**, 53, 2449 – 2452.

[7] Han, M. Y.; Ozyilmaz, B.; Zhang, Y.; Kim, P. Phys. Rev. Lett. **2007**, 98, 206805.

[8] Berger, C.; Song, Z.; Li, X.; Wu, X.; Brown, N.; Naud, C.; Mayou, D.; Li, T.; Sprinkle, M.; Hass, J.; Marchenkov, A. N.; Conrad, E. H.; First, P. N.; de Heer, W. A. Science, **2006**, 312, 1191 – 1196.

[9] de Heer, W. A.; Berger, C.; Wu, X.; First, P. N.; Conrad, A. N.; Li, X.; Li, T.; Sprinkle, M.; Hass, J.; Sadowski, M. L.; Potemski, M.; Martinez, G. Solid State Commun. **2007**, 143, 92 – 100.

[10] Riedl, C.; Starke, U.; Bernhardt, J.; Franke, M.; Heinz, K. Phys. Rev. B, **2007**, 76, 245406.

[11] Ohta, T.; Bostwick, A.; Seyller, T.; Horn, K.; Rotenberg, E. Science, **2006**, 313, 951 – 954.

[12] Emtsev, K.V.; Seyller, Th.; Ley, L.; Tadich, A.; Broekman, L.; Riley, J. D.; Leckey, R.C.G.; Preuss, M. Phys. Rev. B **2006**, 730, 75412.

[13] Gupta, A.; Chen, G.; Joshi, P.; Tadigadapa, S.; Eklund, P.C. Nanolett. **2006**, 6, 2667.

[14] Ferrari, A. C.; Meyer, J. C.; Scardaci, V.; Casiraghi, C.; Lazzeri, M.; Mauri, F.; Piscanec, S.; Jiang, D.; Novoselov, K. S.; Roth, S.; Geim, A. K. Phys. Rev. Lett. **2007**, 97, 189401.

[15] Malard, L. M.; Nilsson, J.; Elias, D. C.; Brant, J. C.; Plentz, F.; Alves, E. S.; Castro Neto, A. H.; Pimenta, M. A. Phys. Rev. B, **2007**, 76, 201401.

[16] Casiraghi, C.; Pisana, S.; Novoselov, K. S.; Geim A. K.; Ferrari A. C. Appl. Phys. Lett. **2007**, 91, 233108.

[17] Graf, D.; Molitor, F.; Ensslin, K.; Stampfer, C.; Jungen, A.; Hierold, C.; Wirtz, L. Nano Lett. **2007**, 7, 238 – 242.

[18] Faugeras, C.; Nerrière, A.; Potemski, M.; Mahmood, A.; Dujardin, E.; Berger, C.; de Heer, W. A. Appl. Phys. Lett. **2008**, 92, 011914.

[19] Ni, Z. H.; Chen, W.; Fan, X. F.; Kuo, J. L.; Yu, T.; Wee, A. T. S.; Shen, Z. X. Phys. Rev. B **2008**, 77, 115416.

[20] Röhrl, J.; Hundhausen, M.; Emtsev, K. V.; Seyller, Th.; Graupner, R.; Ley, L. Appl. Phys. Lett. **2008**, 92, 201918.

[21] van Bommel, A. J.; Crombeen, J. E.; van Tooren, A. Surf. Sci. **1975**, 48, 463 – 472.

[22] Forbeaux, I; Thermlin, J.-M.; Debever, J.-M. Phys. Rev. B **1998**, 58, 16396-16406.

[23] Soubatch, S.; Saddow, S.E.; Rao, S. P.; Lee, W. Y.; Konuma, M.; Starke, U. Mater. Sci. Forum **2005**, 483-485, 761.

[24] Starke, U. in *Silicon Carbide, Recent Major Advances*, edited by W. J. Choyke, H. Matsunami, and G. Pensl (Springer, Berlin, 2004), pp. 281–316.

[25] Riedl, C.; Zakharov, A. A.; Starke, U. Appl. Phys. Lett. **2008**, 93, 033106.

[26] Varchon, F.; Feng, R.; Hass, J.; Li, X.; Ngoc Nguyen, B.; Naud, C.; Mallet, P.; Veuillen, J.-Y.; Berger, C.; Conrad, E. H.; Magaud, L. Phys. Rev. Lett. **2007**, 99, 126805.

[27] Zhou, S. Y.; Gweon, G.-H.; Fedorov, A. V.; First, P. N.; de Heer, W. A.; Lee, D.-H.; Guinea, F.; Castro Neto, A. H.; Lanzara, A. Nat. Mater. **2007**, 6, 770 – 775.

[28] Sharf, T. W.; Singer, I. L. Trib. Lett. **2003**, 14, 137 – 145.

[29] Hass, J.; Varchon, F.; Millán-Otoya, J. E.; Sprinkle, M.; Sharma, N.; de Heer, W. A.; Berger, C.; First, P. N.; Magaud, L.; Conrad, E. H. Phys. Rev. Lett. **2008**, 100, 125504.

[30] Hibino, H.; Kageshima, H.; Maeda, F.; Nagase, M.; Kobayashi, Y.; Yamaguchi, H. Phys. Rev. B **2008**, 77, 075413.

[31] Ohta, T.; Gabaly, F. El.; Bostwick, A; McChesney, J. L.; Emtsev, K. V.; Schmid, A. K.; Seyller, Th.; Horn, K.; Rotenberg, E. New J. Phys. **2008**, 10, 023034.

[32] Pisana, S.; Lazzeri, M.; Casiraghi, C.; Novoselov, K.; Geim, A. K.; Ferrari, A. C.; Mauri, F. Nat. Mater. **2007**, 6, 198 – 201.

[33] Yan, J.; Zhang, Y. B.; Kim, P.; Pinczuk, A. Phys. Rev. Lett. **2007**, 98, 166802.

[34] Stampfer, C.; Molitor, F.; Graf, D.; Ensslin, K.; Jungen, A.; Hierold, C.; Wirtz, L. Appl. Phys. Lett. **2007**, 91, 241907.

[35] Das, A.; Pisana, S.; Chakraborty, B.; Piscanec, S.; Saha, S. K.; Waghmare, U. V.; Novoselov, K. S.; Krishnamurthy, H. R.; Geim, A. K.; Ferrari, A. C.; Sood, A. K. Nat. Nanotechnol. **2008**, 3, 210 – 215.

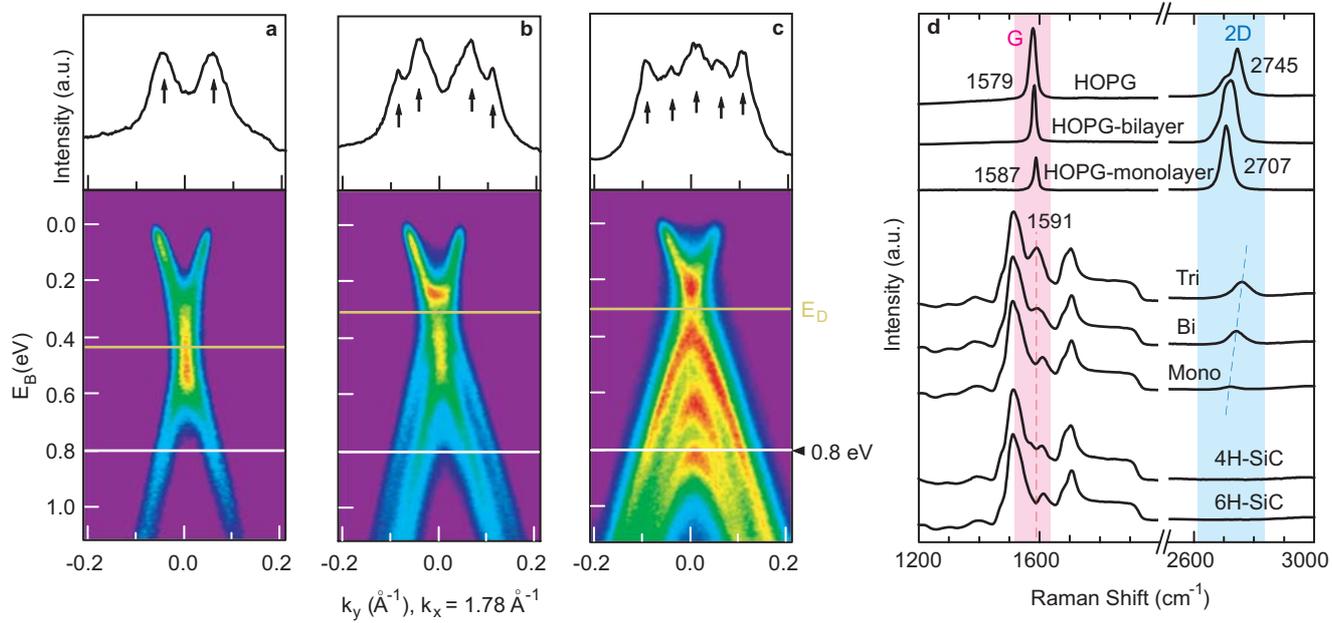

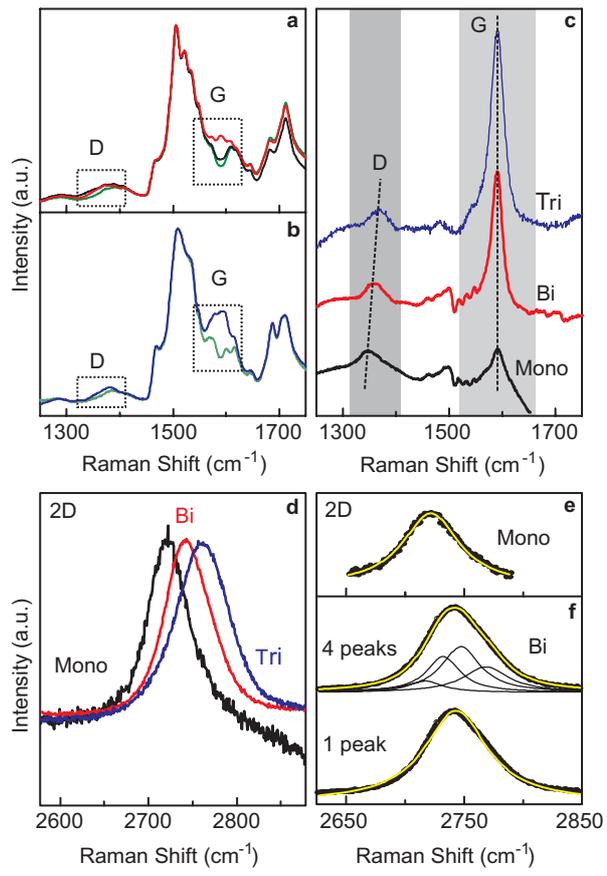

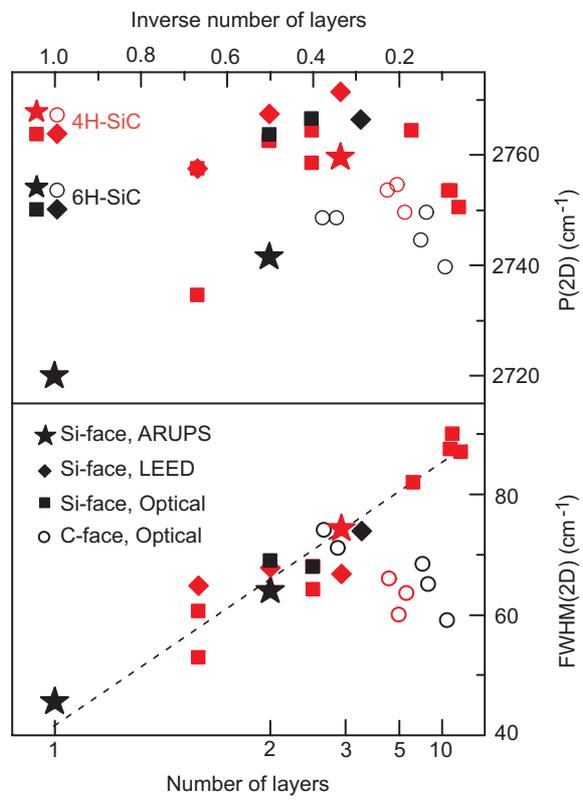

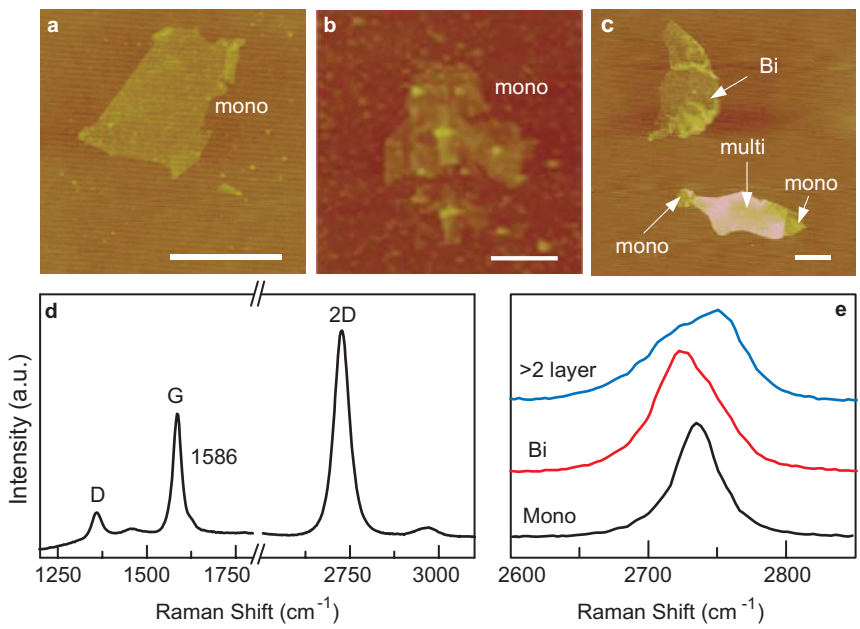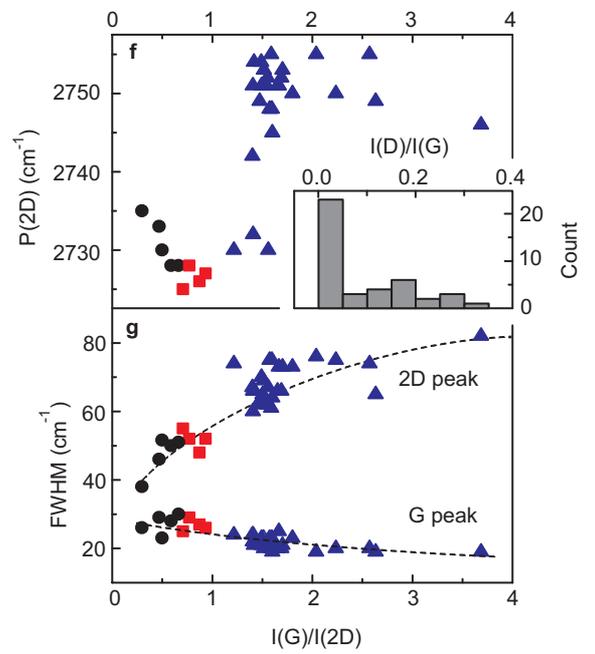

SUPPORTING INFORMATION for

Raman spectra of epitaxial graphene on SiC and of epitaxial graphene transferred to $SiO_2$

*Dong Su Lee, Christian Riedl, Benjamin Krauß, Klaus von Klitzing, Ulrich Starke and Jurgen H. Smet*

A. AFM measurements on epitaxial graphene

**Figure S1: AFM images of (a) 4H-SiC(0001) after $H_2$ etching, (b) mono-, (c) bi- and (d) trilayer graphene grown on SiC. All the scale bars correspond to 5 μm.**

Atomically flat terrace structures are formed on the surface of SiC after the $H_2$ etching process. Typically, the width of the terraces is ~ 2 μm and the step height is ~ 2 nm (for 4H-SiC(0001), cf. Fig. S1a). After graphitization, the samples display morphologies with small terrace structures (cf. Fig. S1b-d) quite different from the initial terraces formed during the $H_2$ etching [1–4]. The usual step height for the small terraces after graphitization corresponds to half of the substrate unit cell, i.e. ~ 7 – 8 Å in the case of mono- and bilayer graphene (6H-SiC), and 5 Å in the case of trilayer graphene (4H-SiC). For monolayer graphene, the initial terrace structures are still visible. For instance, the long single terrace between the white dashed lines marked in Fig. S1b apparently represents the initial terrace structure. In thicker layer graphene, however, the initial terraces are hardly visible.

B. Determination of the number of graphene layers using an attenuation model

**Figure S2: The ratio between the intensity of the strongest SiC substrate Raman peak obtained on graphitized samples and on a hydrogen etched reference sample as a function of the graphene layer thickness. Experimental data points are for mono-, bi- and trilayer samples (mainly discussed in the manuscript) and a 3 to 4 layer thick graphene sample. The error bars mark the maximum and minimum values from a set of 4 or 5 measurements taken at different positions on the sample. The red curve corresponds to an exponential attenuation fit.**

In order to estimate the thickness of epitaxial films without ARUPS or LEED a simple model based on Beer's Law [5, 6] was used. The model assumes that light passing through a film is attenuated exponentially according to exp(-$t/\lambda$), where $t$ is the film thickness and $\lambda$ the penetration depth of the laser light. The SiC peak at ~ 1510 cm$^{-1}$, which has the highest intensity, was used for these purposes (see Fig. 2 in the manuscript). In Fig. S2 we plot this peak intensity as a function of the number of graphene layers for graphene samples, whose layer thicknesses were predetermined from ARUPS or LEED. These intensities were normalized to the intensity of this peak recorded on the bare SiC substrate, $I_{Ref}$. The data points are fitted to $I_{Sample}/I_{Ref}$ = exp(-$2t_0N/\lambda$), where $t_0$ is the stacking distance of the graphene layers and $N$ the number of layers. The factor of 2 is added to take into account the reflection geometry. The fitting parameter amounts to $2t_0/\lambda$ ~ 1/10, and thus $\lambda$ ~ $20t_0$. From the graphite stacking distance $t_0$ = 0.335 nm, we deduce that $\lambda$ is about 6.7 nm. This fit serves to determine the number of layers on samples on which no LEED or ARUPS data were available. For thick graphene (more than 6 layers) the estimated thickness will exhibit a larger error since we extrapolate from data recorded on few layer graphene only. Applying this procedure to the thickest graphene samples used in our studies, we estimated layer numbers between 11 and 14 (the uppermost three red squares in the lower panel of Fig. 3 in the manuscript). X-ray photoelectron spectroscopy (XPS) was carried out on these thicker layers as an additional check. The emission intensity of the core level also follows a similar attenuation model and the XPS analysis led to an estimated ~ 6 – 9 layers instead.

C. Transfer of graphene flakes grown on SiC onto oxidized Si-substrates.

**Figure S3: Optical micrographs of graphene transferred from SiC onto Si-substrates with a thermally grown SiO$_2$-layer. The contrast of the image was enhanced to improve the visibility of the flakes. The red arrows indicate the visible blueish objects on which further Raman measurements were carried out to identify them as graphene or glue residues. The yellow objects are Au markers deposited by conventional e-beam lithography and thermal evaporation. They help in relocating flakes. The spacing between the markers is 80 µm.**

Flakes of epitaxial graphene grown on SiC were transferred onto Si-substrates with a 300 nm thick thermal oxide layer at the top. The conventional adhesive tape technique was used for the exfoliation. As starting material, we used a trilayer sample and a sample with a thickness varying from one to four layers. Both samples were grown on a 4H-SiC(0001) substrate. The yield and the properties of the flakes did not depend on the specific starting material. After attaching several times a small piece of adhesive tape onto the graphitized substrate, graphene flakes were transferred by gently pressing the tape onto the oxidized Si-substrate and subsequently rubbing it with a soft tip of plastic tweezers. The adhesive tape that is commonly used for dicing wafers as well as mechanical exfoliation of graphene flakes from HOPG because it leaves little residue, does not offer sufficient adhesion and hence a tape with a larger glueing force was needed. In most cases, the transferred graphene flakes were ~ 1 $\mu m^2$ or less in size, and thus hardly visible by eye using an optical microscope. Even if they were visible, it was hard to distinguish the graphene flakes from glue residues. For example, Fig. S3 shows optical micrographs made on an oxidized Si-substrate on which epitaxial graphene flakes were transferred. The substrate was cleaned with acetone, isopropanol and DI-water before optical inspection. Four blueish objects are highlighted with red boxes and red arrows. However, only two of them were identified as graphene flakes based on Raman measurements.

D. Spatial Raman map on an epitaxial monolayer of SiC based graphene

**Figure S4: a: Rayleigh image of a monolayer of graphene grown on SiC. Regions with yellow and black colors indicate a high and a low optical reflection signal respectively. The difference between the signals for yellow and black regions is approximately 4 %. The scale bar is 2 μm. The bottom is a cartoon of the graphene morphology. It follows the terrace structure of the SiC substrate. b: The top panel depicts a cross section at the blue line of the optical reflection map of panel a. The bottom panel displays the FWHM(2D) in the Raman spectrum at various locations. The dashed lines mark the position of terrace edges as seen from the optical reflection signal.**

Figure S4**a** displays a Rayleigh map for a monolayer of graphene fabricated on SiC. This map illustrates the typical morphology of such samples [7]. The spatial map resembles the AFM image of the sample (Fig. S1**b**). Raman spectra were recorded along the blue dashed line in 500 nm steps with a laser spot size of ~ 400 nm. Figure S4**b** shows a cross section of the Rayleigh signal of panel **a.** In the bottom panel the width of the 2D peak, i.e. FWHM(2D), is plotted as measured along that line at the same positions. The FWHM(2D) values undulate around 44 – 48 $cm^{-1}$. These are widths characteristic for monolayer graphene. However, near the edges of the terraces the widths are larger by 2 – 3 $cm^{-1}$ than on the terraces themselves. These increased values are still too small to originate from a bilayer region. (We note that also P(2D) is constant around 2723 $cm^{-1}$.) However, since it is expected that at the edges of the terraces the graphene sheet is bent [8], the broadening of the 2D peak may be explained by the curvature at the edge and the accompanied local displacement of carbon atoms in the honeycomb lattice to conform to the terrace structure. This observed variation in the FWHM(2D) at the terrace edges is not sufficient to explain the larger difference in the full FWHM(2D) of about 10 – 20 $cm^{-1}$ between graphene grown on SiC and graphene exfoliated from HOPG. Since the larger broadening for SiC based graphene does not disappear upon transferring the graphene to an oxidized Si substrate, we propose that structural defects such as vacancies to accommodate these terraces or even local displacement of C-atoms due to the interaction between the reconstruction and graphene are responsible for this large difference in the FWHM.

References for Supporting Information

[1] Riedl, C.; Starke, U.; Bernhardt, J.; Franke, M.; Heinz, K. Phys. Rev. B, **2007**, 76, 245406.

[2] Riedl, C.; Zakharov, A. A., Starke, U. Appl. Phys. Lett., accepted.

[3] Hibino, H.; Kageshima, H.; Maeda, F.; Nagase, M.; Kobayashi, Y.; Yamaguchi, H. Phys. Rev. B **2008**, 77, 075413.

[4] Ohta, T.; Gabaly, F. El.; Bostwick, A; McChesney, J. L.; Emtsev, K. V.; Schmid, A. K.; Seyller, Th.; Horn, K.; Rotenberg, E. New J. Phys. **2008**, 10, 023034.

[5] Ni, Z. H.; Chen, W.; Fan, X. F.; Kuo, J. L.; Yu, T.; Wee, A. T. S.; Shen, Z. X. Phys. Rev. B **2008**, 77, 115416.

[6] Sharf, T. W.; Singer, I. L. Trib. Lett. **2003**, 14, 137 – 145.

[7] Casiraghi, C.; Hartschuh, A.; Lidorikis, E.; Qian, H.; Harutyunyan, H.; Gokus, T.; Novoselov, K. S.; Ferrari, A. C. Nanolett, **2007**, 7, 2711 – 2717.

[8] Seyller, Th.; Emtsev, K. V.; Gao, K.; Speck, F.; Ley, L.; Tadich, A. Broekman, L.; Riley, J. D.; Leckey, R. C. G.; Rader, O.; Varykhalov, A.; Shikin, A. M. Surf. Sci. **2006**, 600, 3906 – 3911.

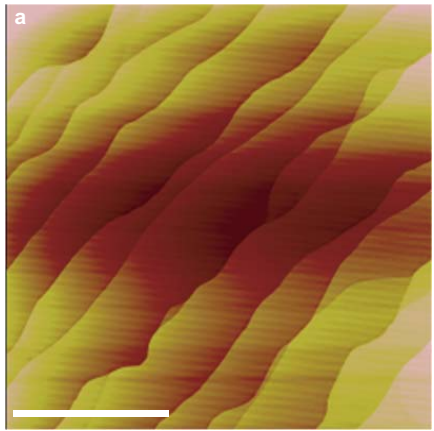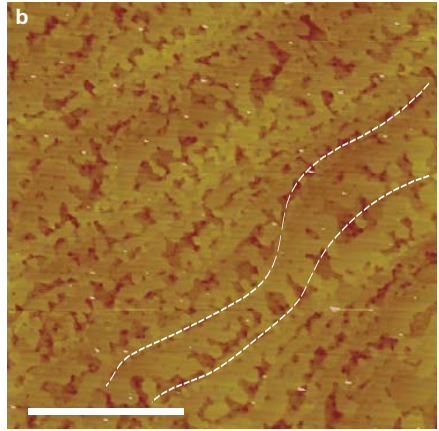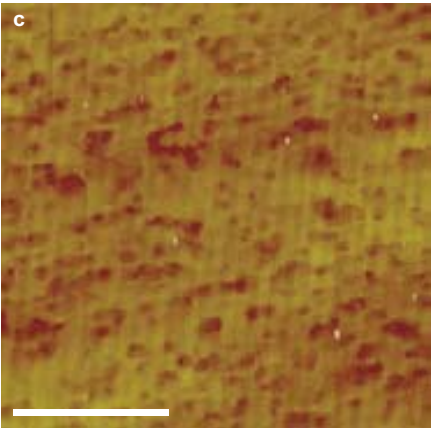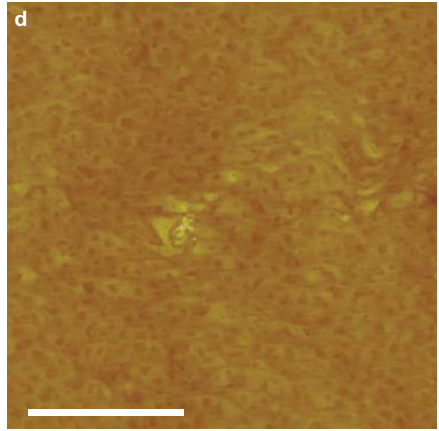

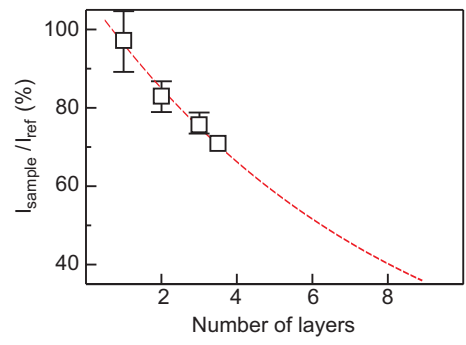

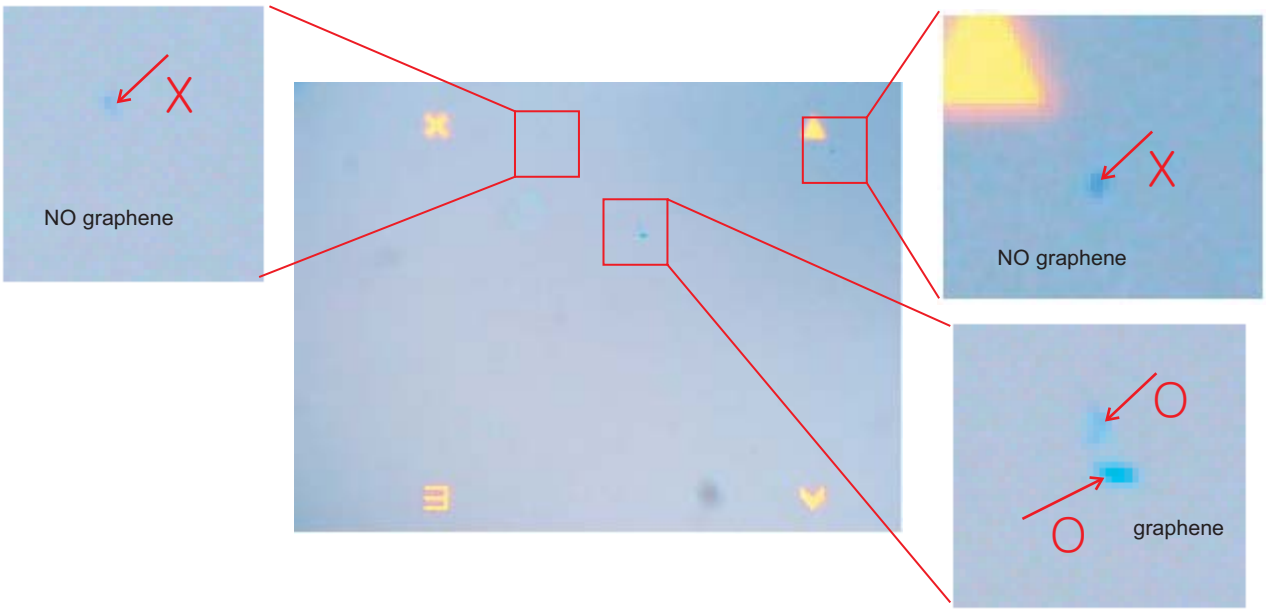

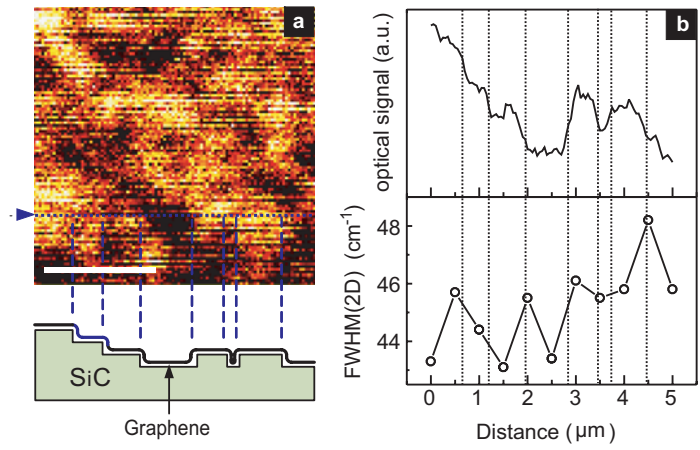